\documentclass[aip,jap,amsmath,amssymb,reprint,floatfix]{revtex4-1}

\usepackage{graphicx}
\usepackage{mathptmx}

\makeatletter
\def\@email#1#2{%
 \endgroup
 \patchcmd{\titleblock@produce}
  {\frontmatter@RRAPformat}
  {\frontmatter@RRAPformat{\produce@RRAP{*#1\href{mailto:#2}{#2}}}\frontmatter@RRAPformat}
  {}{}
}%
\makeatother
\begin{document}

\preprint{AIP/123-QED}

\title{Statistical Disorder in MBE-Grown AlGaAs/GaAs Superlattices for Quantum Bragg Mirrors using Synchrotron X-ray Diffraction}

\author{M. T. Souza}
\author{G. M. Penello}%
\affiliation{Institute of Physics, University of S\~ao Paulo, S\~ao Paulo, SP, Brazil}

\author{Guilherme A. Calligaris}
\affiliation{Brazilian Synchrotron Light Laboratory - SIRIUS/CNPEM, Campinas, SP, Brazil}

\author{S\'ergio L. Morelh\~ao}
\affiliation{Institute of Physics, University of S\~ao Paulo, S\~ao Paulo, SP, Brazil}


\date{\today}

\begin{abstract}

AlGaAs/GaAs superlattices grown by Molecular Beam Epitaxy (MBE) are foundational for advanced optoelectronic devices, including Quantum Bragg Mirror (QBM) infrared detectors. The performance of these devices critically depends on achieving near-perfect periodicity and abrupt interfaces, however, intrinsic statistical fluctuations during MBE growth introduce nanoscale structural disorder that can degrade device efficiency. In this study, we present a comprehensive methodology for quantifying this disorder in a 203-layer QBM device. High-resolution structural characterization was performed using high-energy (25\,keV) synchrotron X-ray diffraction. By coupling a recursive dynamical diffraction formalism with an ensemble simulated annealing refinement, we extracted statistically robust, layer-by-layer thickness profiles. Our analysis reveals highly systematic, material-specific deviations from the nominal design: all AlGaAs barrier layers were consistently thinner than nominal by 0.3-0.6\,nm. Furthermore, the sequential thickness profile successfully identified  a significant 35\,nm deficit in the final macroscopic top contact layer and a 40\,nm deficit in the first GaAs layer. Achieving a statistical precision of $\pm 0.2$ to $0.6$\,nm (approximately 1-2 atomic monolayers), this non-destructive diagnostic approach provides directly actionable feedback for MBE flux calibration protocols.

\end{abstract}

\maketitle

\section{Introduction}

AlGaAs/GaAs heterostructures occupy a prominent position among the semiconductor material systems that underpin infrared detection technologies \cite{fukuda1998optical,ladugin2019advanced}, applications that play an increasingly important role in areas such as thermal imaging, remote sensing, environmental imaging and astrophysical instrumentation \cite{razeghi2010technology}. AlGaAs/GaAs heterostructures, with their near-perfect lattice match to the GaAs substrate, which eliminates
strain-related defects that would otherwise limit device performance, and their
well-understood band alignment that offers considerable flexibility in band-gap engineering, show continued reliability for use in the development of such technologies\cite{fukuda1998optical,ladugin2019advanced}.

 Beyond the established architectures, AlGaAs/GaAs heterostructures have also fostered continuous interest for use in novel quantum devices such as with Quantum Bragg Mirror (QBM) infrared detectors \cite{penello2024tailoring,penello2023gaas,Appas:22,Baboux:23,sharma2024theoretical,tsang1986gaas}, with such devices requiring the controlled formation of ultra-thin layers with high compositional uniformity and
reproducible periodicity \cite{ladugin2019advanced}. Though these structural requirements can typically be met through crystal growth by Molecular Beam Epitaxy (MBE)\cite{manfra2014molecular,kikkawa1990growth}, the statistical fluctuations in layer thickness and
composition that are inherent to the MBE process introduce a degree of structural
disorder that can degrade device efficiency or reduce fabrication yield \cite{chand1989mbe,morkocc1982influence,chand1993growth,radulescu1988influence}.

The ability to quantify nanoscale structural disorder is, therefore,
essential both for the optimization of MBE growth protocols and for accurate modeling of
device performance. Structural characterization by X-ray diffraction and scattering has
been well established as a tool for probing layer thickness, composition, and
interface quality at the nanometer and sub-nanometer scale \cite{penacchio2022statistical,segmuller1991characterization,segmuller1989x,morelhao2016computer,als2011elements,sasaki1996using,takahasi2018situ},
making it a natural choice for this type of analysis.

In this study we present a systematic procedure for extracting quantitative structural  
parameters from synchrotron X-ray diffraction data through a detailed investigation of a 20-period AlGaAs/GaAs
superlattice designed for QBM applications. High-energy (25\,keV) diffraction
measurements were performed at the Sirius light source, providing high angular resolution. We employed a dynamical diffraction model,
refined with the use of a simulated annealing algorithm, to extract the statistical parameters
describing layer-thickness and composition. The resulting quantitative
characterization of nanoscale disorder provides direct feedback for MBE process
refinement27,6 and a foundation for more accurate prediction of QBM device performance.

\section{Methods}

The QBM detector sample in this work was designed for use in the far-infrared range, and is composed of an active region with five coupled quantum-wells in an asymmetric structural configuration. The aim of this structure is of optimizing the device's photovoltaic configuration mode. This active region was repeated 20 times to form the complete device, which comprises 203 layers in total, including quantum wells, quantum barriers, and contact layers. \cite{penello2025structural}

\begin{figure}                 
    \includegraphics[width=0.9\columnwidth]{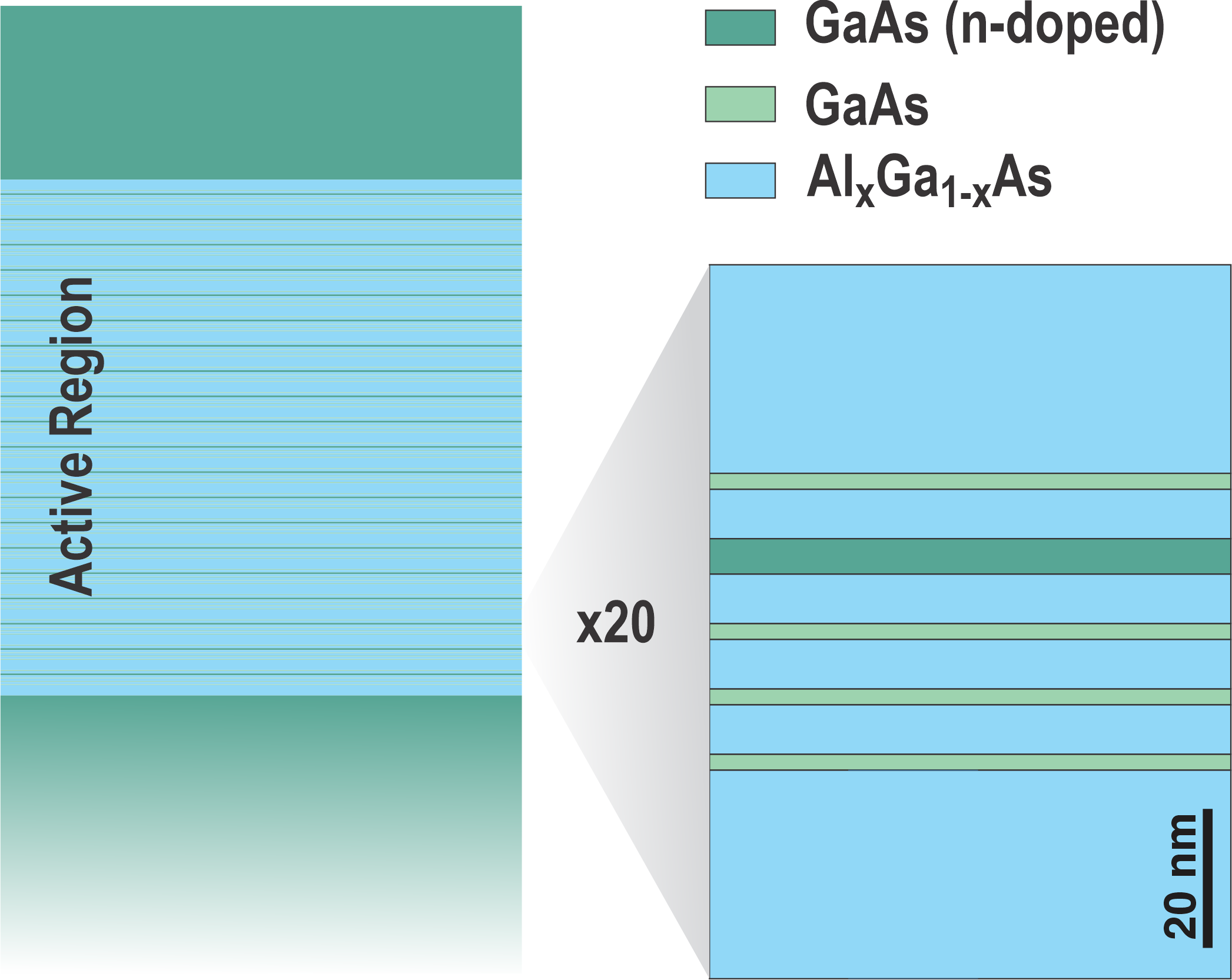}
    \caption{Schematic diagram of the nominal device heterostructure (left) and the detailed layer sequence of a single 20× repeating unit within the active region (right), consisting of alternating GaAs (undoped and n-doped) and Al$_x$Ga$_{1-x}$As layers.}
    \label{fig:nominalstructure}
\end{figure}

The QBM detector sample was grown by molecular beam epitaxy (MBE) on an epi-ready GaAs(001) substrate. Prior to growth, out-gassing and oxide removal were done at 600\,\textdegree C and 580\,\textdegree C respectively. Reflection high-energy electron diffraction (RHEED) was employed to calibrate in situ the composition and growth rate of each constituent layer, ensuring the compositional and thickness precision required by the multilayer design of the device. The thin layers of the active region were grown at 580\,\textdegree C at a rate of 0.28\,nm/s and each period of the active region included a main quantum well Si-doped with $n = 1 \times 10^{18}\,{\rm cm}^{-3}$ to populate the ground state. To obtain good Ohmic contacts during processing, the top and bottom GaAs contact layers were doped with the same Si concentration of $n = 1 \times 10^{18}\,{\rm cm}^{-3}$. \cite{penello2025structural}

Synchrotron X-ray diffraction (XRD) measurements were performed at the EMA beamline of the Sirius light source (LNLS/CNPEM, Brazil). The experiment utilized a high-flux, low-divergence beam at an incident energy of 25\,keV, and diffraction patterns were collected using a Dectris Pilatus 300K area detector.

\section{X-ray diffraction simulation and structural parameter refinement}

\subsection{The recursive dynamical diffraction formalism}

 For the X-ray diffraction simulation developed for this work, the Darwin–Prins dynamical diffraction theory was used as its theoretical basis. This is a formalism developed by considering the multiple scattering events experienced by X-ray waves propagating through a periodic crystalline medium. \cite{morelhao2016computer,als2011elements}
 
 Let X and Y be two monolayers (MLs) such that their reflection and transmission coefficients are $R_{X,Y}$ and $T_{X,Y}$.Then the coefficients of the combined X:Y double layer (with X on top of Y) can be calculated according to Morelh\~ao et al.\cite{morelhao2017nanoscale}:

\begin{equation}\label{eq1}   
\begin{split}
    R_{XY} &= R_X + R_Y \frac{T_X^2 \cdot \exp(2i\varphi)}{1 - \bar{R}_{X} R_Y\cdot \exp(2i\varphi)}, \\
\bar{R}_{{XY}} &= \bar{R}_{Y} + \bar{R}_{X} \frac{T_Y^2 \cdot \exp(2i\varphi)}{1 - \bar{R}_{X} R_Y\cdot \exp(2i\varphi)}, \\
T_{XY} &= \frac{T_X T_Y\cdot \exp(2i\varphi)}{1 - \bar{R}_{X} R_Y\cdot \exp(2i\varphi)}
\end{split}
\end{equation}
where R and $\bar{R}$ denote the forward and backward reflection amplitudes, respectively, while T represents the transmission amplitude. $\varphi = -\frac{1}{2} Qd$ is the phase delay accumulated each time an X-ray wave of wavelength $\lambda$ traverses the interlayer distance $d$ between the MLs X and Y, and $Q = (4\pi /\lambda)\sin\theta$ is the modulus of the scattering
vector perpendicular to the MLs for an incident angle $\theta$.

To incorporate a third ML Z at the bottom of the double layer X:Y, Eq.( \ref{eq1}) can be used recursively by replacing the coefficients of the combined double layer X:Y in place of X, and those of monolayer Z in place of Y, with $d$ now equal to the interlayer distance between Y and Z. This results in the reflection and transmission coefficients of the triple layer X:Y:Z. The procedure can be extended recursively to an arbitrary number of monolayers, as illustrated below:

\begin{equation}\label{eq2}
    \begin{split}
    (X = A,Y= B)\rightarrow&  R_{AB},\\
    (X = AB,Y = C)\rightarrow& R_{ABC},\\
    (X = ABC,Y = A)\rightarrow& R_{ABCA},\\
    (X = ABCA\dots,Y =N)\rightarrow& R_{ABC\dots N}.
    \end{split}
\end{equation}

These recursive calculations described by Eqs.\,(\ref{eq1}) and \,(\ref{eq2}) have been demonstrated to provide the dynamical diffraction solution \cite{morelhao2017nanoscale} in specular reflection geometry, in which refraction, rescattering, and photoelectric absorption are taken into account.

\subsection{The Simulated Annealing algorithm}

To determine the structural parameters that best reproduce the experimentally measured X-ray reflectivity curve, we employed a simulated annealing algorithm.

The simulated annealing algorithm is a probabilistic optimization technique designed to locate the global minimum of a cost function that may possess multiple local minima. This method gets its name from an analogy with the physical annealing process, in which a crystalline solid is heated to a temperature sufficient to allow extensive atomic rearrangement and then gradually cooled, so that the final structure is \textit{frozen} in the lowest accessible energy configuration.\cite{bertsimas1993simulated,rutenbar2002simulated} 

In the context of this study, we call a combinatorial optimization problem one that consists of finding an optimal configuration of relevant parameters $\bar{X}=(X_1,X_2,\dots,X_N)$ that minimizes a given function $f(\bar{X)}$, usually referred to as the cost or objective function, which serves as a quantitative metric for the quality of a specific solution. An iterative approach, applying small random perturbations and accepting only cost-reducing moves until no further improvement is possible, can be prone to becoming trapped in local minima rather than locating the global minimum, and because of that a simulated annealing approach extends this strategy by incorporating a mechanism that allows escape from local minima. For that to be possible, cost-increasing perturbations (moves) are carefully allowed. With this process, perturbations can displace a configuration of parameters out of a local minimum by selectively accepting worse solutions in a controlled manner, thereby enabling the algorithm to locate a more promising descent trajectory toward the global minimum. \cite{rutenbar2002simulated}

In practice, random perturbations are applied to the current parameter configuration and
evaluated by the cost function, with lower-cost solutions being always accepted while
higher-cost solutions may still be accepted with a non-zero probability. In physical
systems, the probability of a transition to a higher energy state is governed by the
system's temperature: higher temperatures correspond to a greater likelihood of such
transitions. To emulate this behavior, the so-called Metropolis algorithm \cite{metropolis1953equation} assigns an acceptance probability based on the Boltzmann
distribution for a change $\Delta C$ in the cost function at temperature $T$:
\begin{equation}\label{eq3}
    P_{\text{acc}} = e^{-\Delta C / T}.
\end{equation}

The algorithm proceeds by generating a random number $R$ uniformly distributed on the
interval $[0, 1]$ and comparing it against $P_{\text{acc}}$; a move is accepted only if
$R < P_{\text{acc}}$. For cost-reducing moves ($\Delta C < 0$), $P_{\text{acc}} > 1$ and
the move is always accepted. For cost-increasing moves ($\Delta C > 0$), $P_{\text{acc}}
< 1$ and acceptance becomes exponentially less likely as the cost increase grows larger.
This mechanism allows the algorithm to escape local minima by occasionally accepting
worse solutions, while still favoring improvements. As the temperature $T$ is
progressively lowered over the course of the optimization, the probability of accepting
cost-increasing moves decreases, effectively emulating a physical system approaching
thermal equilibrium through the annealing process.
\cite{rutenbar2002simulated}

The simulated annealing algorithm can be applied to arbitrary combinatorial optimization
problems, but in the context of this study it was used for the identification of the crystal structure
parameters that best reproduce a given experimental reflectivity curve. The temperature
$T$ serves as a control parameter, moderated through a cooling schedule: a sequence of
decreasing temperatures designed to progressively reduce the acceptance of cost-increasing
moves as the optimization proceeds \cite{rutenbar2002simulated}.

\subsection{Structural parameter refinement of the QBM device}

The known nominal crystal structure of the QBM device, comprising 203 layers, was used
as the initial parameter configuration for the simulated annealing refinement.
We then computed the X-ray reflectivity curve by summing the reflection and transmission
coefficients of successive layers through the layer stack, obtained via the recursive
method described by Eqs.~(\ref{eq1}) and (\ref{eq2}). The simulated curve was normalized to match the maximum peak amplitude of the experimental X-ray reflectivity
data, ensuring a consistent intensity scale for comparison.

Following this, a cost function quantifying the discrepancy between the experimental and
simulated curves in logarithmic space was defined as:
\begin{equation} \label{eq4}
    \chi^2(\mathbf{NL}) = \sum_{i=1}^{N} \left[ \log_{10} \left(
    \frac{\max \{ I_i^{\mathrm{exp}}, \epsilon \}}
         {\max \{ \mathcal{I}_{\mathrm{sim}}(Q_i^{\mathrm{exp}};\, \mathbf{NL}),
         \epsilon \}} \right) \right]^2
\end{equation}
where $\mathbf{NL}$ denotes the vector of 203 layer thickness
parameters, $(Q_i^{\text{exp}}, I_i^{\text{exp}})$ are the experimental momentum
transfer values and corresponding intensities, $\epsilon = 10^{-12}$ is a small
regularization constant introduced to avoid undefined logarithmic arguments and
$\mathcal{I}_{\text{sim}}(Q;\,\mathbf{NL})$ denotes the simulated intensity, evaluated
at the experimental $Q$-values. The use of logarithmic space is so the cost function is ensured to have weights deviations across the full dynamic range of the reflectivity curve
rather than being dominated by the strongest peaks.

With the cost function defined, the simulated annealing algorithm was employed to search for layer configurations that better reproduce the experimental data. The approach is to perform layer thickness variations from the nominal 203-layer parameter set, with these random perturbations of the layer thicknesses iteratively applied and evaluated. This process allows the algorithm to explore the parameter space and identify structures with lower cost, when one does exist.

Given the non-deterministic nature of the simulated annealing algorithm, the
optimization procedure was repeated 30 times, yielding 30 independent structural
solutions of the studied parameters. The mean and standard deviation of each layer thickness were
then computed across these solutions, providing a statistically robust estimate of the
device structure together with quantified uncertainties for each refined parameter.

\section{Results and Discussion}

\subsection{Characterization and simulation results}

\begin{figure*}[htb]                       
    \centering
    \includegraphics[width=\textwidth]{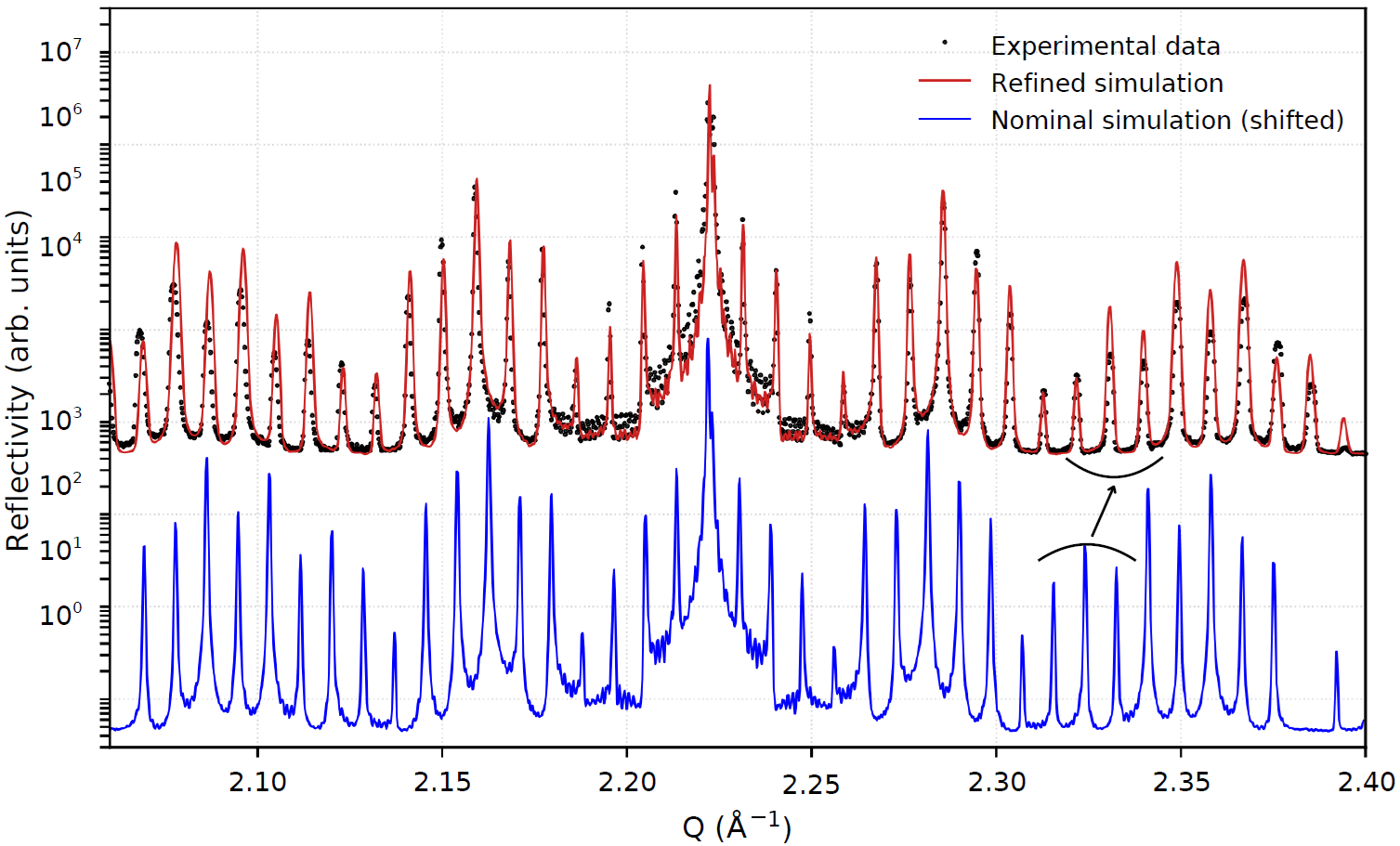}
    \caption{%
        Comparison of experimental and simulated X‑ray reflectivity curve for
        the AlGaAs/GaAs superlattice. Refined parameters simulation (red),
        $\chi^2 = 1.49$ and nominal parameters simulation (blue), $\chi^2 = 4.98$ around the 002 peak. 
    }
    \label{fig:Figure 1}
\end{figure*}

The experimental X-ray reflectivity curve collected at the EMA beamline of the Sirius
synchrotron light source is presented in Fig.~\ref{fig:Figure 1}, together with the
simulated curve obtained from the ensemble-averaged refined structural parameters and the simulation of the nominal structural parameters. Note that the presented Q-scan focuses specifically on the (002) reflection, limiting the evaluated Q-range.

The simulation reproduces the experimental data with good fidelity across the measured$ Q$-range of approximately 2.0--2.4\,\AA$^{-1}$, capturing the position and relative
intensity of the main Bragg reflection near $Q \approx 2.22$\,\AA$^{-1}$ as well as the
superlattice satellite peaks positions on both sides. The overall agreement is reflected in a
cost function value of $\chi^2_{sim}=1.49$, representing a substantial improvement
over the nominal structure simulation, which yields $\chi^2_{nom} = 4.98 $ as the peaks positions, specially those of the satellite peaks, of the nominal structure simulation are badly misaligned with the experimental data. This improvement confirms that the ensemble simulated annealing refinement is recovering
physically meaningful structural information rather than exploring a degenerate region
of the fitting landscape. Existing broadening of the half-width of satellite peaks  in the experimental data when compared to the simulation can be attributed to the existence of interface inhomogeneities and/or broadening \cite{PhysRevB.33.5565}.

The complete refined thickness profile of the active region is shown on Fig.~\ref{fig:Figure 3} a):

\begin{figure*}[ht]                       
    \centering
    \includegraphics[width=0.95\textwidth]{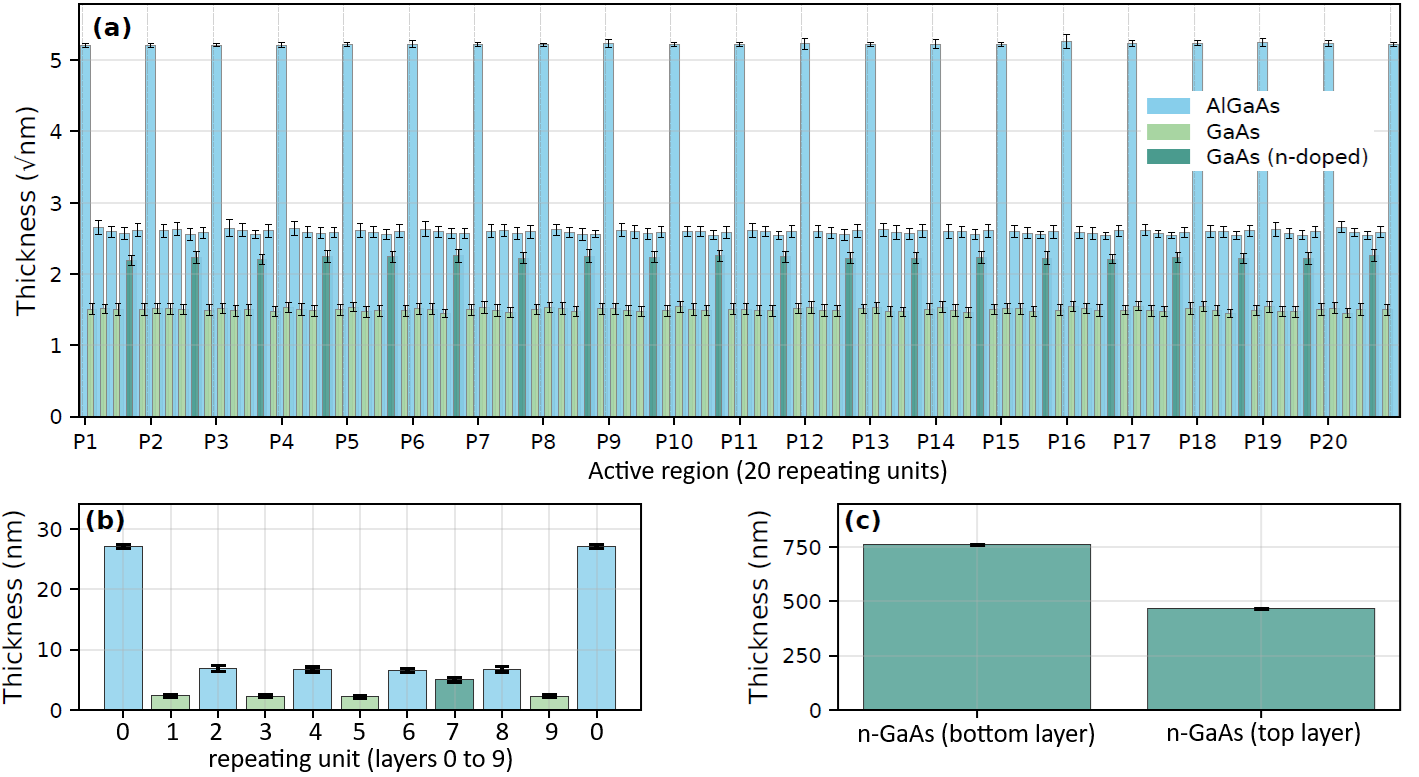}
    \caption{%
    (a) Thickness profile within the active region. (b) Average periodic
structure. (c) Thickness of the top and bottom layers}
    \label{fig:Figure 3}
\end{figure*}

The 10-layer unit cell can clearly be seen repeated with high regularity across all 20 periods, further indicating that the MBE growth process maintained stable conditions throughout the deposition. The tall peak in the beginning of each period, corresponding to the 30\,nm nominal AlGaAs contact barrier, is the most prominent feature of the profile and seems to be consistently reproduced at approximately 27.3 \,nm across periods.

The high consistency visible in Fig.~\ref{fig:Figure 3} a) in every period of the structure has an important implication for the interpretations of the refined parameters: the standard deviations reported for each layer are caused by small but genuine stochastic fluctuations around a reproducible mean. This can be seen as evidence to support the interpretation that the primary structural deviations from the nominal design are systematic in nature.

The refined layer thicknesses, averaged over all 20 periods for each of the 10 layer types in the active region, are presented in Fig.~\ref{fig:Figure 3} b), together with the thicknesses of the first GaAs and contact layers in ~\ref{fig:Figure 3} c). The standard deviations across the 20 periods are approximately 0.2 \,nm for the thin GaAs, 0.4 \,nm for the main GaAs quantum well and around 0.5-0.6 \,nm for the AlGaAs barriers. These values correspond to approximately 1-2 monolayers of both GaAs and AlGaAs in [001], and represent the stochastic period-to-period variability of the MBE growth process under the conditions used, constituting a quantitative characterization of the intrinsic growth disorder of the structure. Though these standard deviations represent only the lower bound of the true structural uncertainty, as they do not account for model uncertainty arising from the assumptions of the diffraction formalism or of the finite number of simulated annealing runs, the fact that 1-2 monolayer uncertainty can explain the X-ray diffraction experimental data indicates good growth quality in the sample device.

However, it is a notable fact that all AlGaAs layers are shown to be consistently thinner than the nominal structure, with deviations of 0.3-0.6 \,nm for the nominally 7.1 \,nm barrier layers and 2.7 \,nm  deviations for the nominal 30 \,nm contact barrier. Though still within the uncertainty, the systematic nature of these deviations, consistent in sign across all layer positions of the same material type and highly reproducible across all 20 periods, can be seen as evidence of an important pattern interpretation for MBE growth physics. Because the Al and Ga fluxes and shutter timing are calibrated independently with AlGaAs/GaAs crystal growth with MBE, a systematic thinning of all AlGaAs layers is consistent with a slight over-estimate of the AlGaAs growth rate in the RHEED calibration, further supported by the observation that the fractional deviation is larger for the thicker AlGaAs contact barrier than for the thinner active region barriers.

The refined thicknesses of the first GaAs and top contact layer  are shown separately in Fig.~\ref{fig:Figure 3} c). Both are shown to be considerably thinner than their nominal values, with the first GaAs layer close to 40 \,nm thinner and the top contact layer close to 35 \,nm thinner, which represent statistically significant deviations, at more than 3 standard deviations of difference from the nominal value. 

This finding has practical implications for device processing: a top contact layer that
is 35\,nm thinner than designed affects the etch depth required for mesa isolation and
the contact resistance of the processed device. The ability of the present methodology
to resolve this deviation independently from the active region parameters demonstrates
its value as a comprehensive structural characterization tool for complex multilayer
devices.

model.

\subsection{Implications for QBM device performance and MBE process refinement}

The systematic layer-type-specific deviations identified in this study have two
distinct sets of implications: one for the performance of the present QBM device, and
one for the optimization of future MBE growth runs.

Regarding device performance, the systematic thinning of the
AlGaAs barriers may affect the tunnel coupling between adjacent
quantum wells, which is a design-critical parameter for the photovoltaic operating mode
of this device. A complete assessment of the performance implications would require
simulating the electronic structure and optical response of the refined device geometry, but having each layer's characterization and its uncertainty allows for an estimate to be made regarding the range for optical transition in the device and how the growth quality affects it.

As for the implications for the MBE process refinement, with the systematic nature of the deviations there is directly actionable feedback. The consistent AlGaAs thinning across all barrier
positions and all 20 periods indicates a reproducible growth rate offset that can be
corrected in subsequent growth runs by adjusting the Al flux or AlGaAs shutter timing
by a calculated amount. The methodology presented in this study is well-suited to verify the effectiveness of
such corrections in subsequent growth runs, providing a quantitative feedback loop
between structural characterization and growth process optimization.

\subsection{Sensitivity to the compositional parameters}

Within the framework being studied in this work, because the composition is not originally included as a free parameter in the cost function on Eq.~\ref{eq4}, the layer thickness estimates obtained by the refinement depend implicitly on the assumed Al concentration used in each simulated reflection coefficient.

To test this dependency, the full ensemble refinement procedure with the Al fraction fixed at an value of 20 \% (rather than the nominal 15\% used in growth) was repeated. The resulting thickness estimates for each individual layer as well as their blocked position thickness agreed with the original refinement within one standard deviation, which indicates that the distinct compositional profiles of each simulation do not produce statistically robust fluctuations of the recovered thickness parameters and suggest that the recovered thickness profile is robust to reasonable errors in the assumed composition.

The reason for this limitation is a reflection on the different manner in which the two parameters impact the diffraction pattern being simulated: layer thickness directly determines the phase accumulated across each layer and therefore controls the angular positions of the main and satellite reflections, whereas Al concentration primarily affects the refractive index contrast between layers and consequently the relative reflected amplitude, with a comparatively minor effect on peak position. Because the cost function in Eq.~\ref{eq4} is being evaluated with basis on the logarithmical intensities over the full Q-range that was measured, this procedure is heavily weighted towards reproducing peak positions and therefore the optimization preferentially resolves thickness over composition.

The present methodology should therefore be understood as a robust tool for layer-thickness characterization in cases where composition is independently known (e.g., from growth calibration), rather than as a joint thickness-composition determination technique, and expanding it to independently constrain composition will require further research.

\section{Conclusion}

Quantifying nanoscale disorder in complex epitaxial heterostructures presents a fundamental challenge: traditional rocking-curve analysis lacks the parameter resolution to distinguish systematic growth drift from stochastic layer fluctuations in devices comprising hundreds of layers. By combining high-energy synchrotron X-ray diffraction with recursive dynamical scattering theory and simulated annealing optimization, we have extracted statistically robust structural parameters for a 20-period AlGaAs/GaAs superlattice designed for quantum Bragg mirror infrared detection. This methodology enabled independent refinement of 203 layer thicknesses, yielding a significant improvement in goodness-of-fit compared to the nominal design and providing quantitative metrics for both systematic deviations and intrinsic growth disorder.

The analysis reveals distinct structural deviations dominated by systematic offsets: all AlGaAs barriers exhibit reproducible thinning consistent with overestimation of the AlGaAs growth rate during RHEED calibration. Statistical fluctuations about the mean values for each layer thickness remain small, with standard deviations of $\sigma_{GaAs} \approx 0.2-0.4$\,nm and $\sigma_{AlGaAs} \approx0.4-  0.5$\,nm (1–2 monolayers), indicating that MBE growth conditions remained highly stable across the 20-period stack. Notably, the methodology independently resolves a 35\,nm deficit in the top GaAs contact layer and a 40\,nm deficit in the first GaAs layer, deviations normally invisible to conventional characterization but that may prove critical for subsequent device processing.

These findings provide actionable feedback for MBE protocol refinement. The consistent sign and magnitude of the AlGaAs thinning indicate that adjusting shutter timing or Al cell temperature by calculable amounts will bring subsequent growths into closer agreement with design specifications.

The quantitative disorder parameters furthermore enable statistically linking fundamental growth phenomena to predicted optoelectronic performance.

Beyond AlGaAs/GaAs superlattices, this approach generalizes to any high-periodicity epitaxial system where disorder governs functionality, including InGaN/GaN light emitters, InAs/GaSb type-II superlattices for infrared detection, and topological insulator heterostructures. The ability to resolve monolayer-scale fluctuations across hundreds of layers establishes a quantitative bridge between growth kinetics and device physics. 

\begin{acknowledgments}
We acknowledge the SIRIUS Light Source of the Brazilian Synchrotron Light Laboratory (LNLS), part of the Brazilian Centre for Research in Energy and Materials (CNPEM), a private non-profit organization under the supervision of the
Brazilian Ministry for Science, Technology and Innovations (MCTI). The EMA beamline staff are acknowledged for their assistance during the experiments under proposal No.
20251057. This work was supported by FAPESP under Grant No. 2023/10775-1. MTS acknowledges financial support from the Unified Scholarship Program of the University of S\~ao Paulo (PUB/USP), project 290 (Edital 2025-2026).
\end{acknowledgments}

\section*{References}
\bibliography{QBM_manuscript}
\end{document}